\documentclass[11pt,a4paper]{llncs}
\usepackage{url}
\usepackage{graphicx}
\usepackage{dcolumn}
\usepackage{rotating}
\usepackage{bm}
\usepackage{verbatim}
\usepackage{amsfonts}
\usepackage{listings}

\title{Bounding Run-Times of Local Adiabatic Algorithms}
\begin{document}
\author{M. V. Panduranga Rao}
\institute{
Department of Computer Science and Automation\\
Indian Institute of Science\\
Bangalore\\
India.\\
pandurang@csa.iisc.ernet.in}

\maketitle

\begin{abstract}
A common trick for designing faster quantum adiabatic algorithms is to
apply the adiabaticity condition locally at every instant.
However it is often difficult to determine the instantaneous gap between
the lowest two eigenvalues, which is an essential ingredient in the adiabaticity 
condition.
In this paper we present a simple linear algebraic technique for 
obtaining a lower bound on the instantaneous gap even in such a situation.
As an illustration,
we investigate the adiabatic unordered search of van Dam et al.~\cite{DamVazirani}
and Roland and Cerf~\cite{SearchLocalAdiabatic} when the non-zero entries of the diagonal final Hamiltonian 
are perturbed by a polynomial (in $\log N$, where $N$ is the length of the unordered list) amount.
We use our technique to derive a bound on the running time of a
local adiabatic schedule in terms of the minimum gap between the lowest two
eigenvalues.
\end{abstract}
\section{Introduction}

Adiabatic Quantum Computation (AQC) has attracted a lot of interest in recent times.
First introduced by Farhi et al.~\cite{Adsip}, this paradigm of computing makes
use of the adiabatic theorem of quantum mechanics. 
Informally, the adiabatic
theorem says that if a physical system is in the ground state
of an initial Hamiltonian that  evolves ``slowly enough"
to a final Hamiltonian, with a non-zero gap between 
the ground state and the first excited state of the Hamiltonian at all 
times, then the probability that the system ends up in the ground state of the final
Hamiltonian approaches unity as the total time of evolution tends to infinity.
This fact is used for solving computational problems as follows. To begin with, the system 
is in the ground state of a suitable Hamiltonian. This initial Hamiltonian
is slowly evolved to a final Hamiltonian whose ground state represents
the solution to the problem. If the running time required for a high probability
of reaching the ground state of the final Hamiltonian is at most polynomial in 
the size of the input, we have an efficient AQC algorithm for the problem.
The maximum rate at which the Hamiltonian can evolve at any instant 
without violating the adiabaticity condition depends inversely on square of the 
gap between the two instantaneous lowest eigenvalues.
In many cases, it is difficult to estimate the gap at every instant of the 
evolution. Then, we strike a compromise by imposing a constant \emph{global} 
delay schedule determined conservatively by the minimum gap over the entire
interval. However, if we do have an estimate of this gap for the entire duration,
we can apply the adiabaticity condition \emph{locally} and speed up the rate
of evolution wherever possible.

In the early days of AQC, efforts were largely focused on adiabatic
optimization algorithms. Given a function $\phi:\{0,1\}^n\rightarrow\mathbb{R}$,
the problem is to find an $x\in\{0,1\}^n$ that minimizes $\phi$. 

van Dam et al.~\cite{DamVazirani} and Roland and Cerf~\cite{SearchLocalAdiabatic}  demonstrated
the  quantum nature of AQC by showing a quadratic speed-up for
unordered search, matching Grover's discrete algorithm~\cite{Grover}.

However, any adiabatic quantum computer can be 
efficiently simulated by a standard quantum computer~\cite{DamVazirani}.
Applications of AQC techniques to small sized random instances of several NP 
complete problems like EXACT-COVER~\cite{Farhi2,Farhi1}, 
CLIQUES~\cite{Clique} and SAT~\cite{Farhi2,Farhi1,Adsip} have been  explored with some success.
Aharonov et al.~\cite{Aharonov} generalized this model by considering local\footnote{A Hamiltonian is said to be local if it involves interactions between only a constant number of particles.} final Hamiltonians
instead of only diagonal ones and showed that this generalized model can 
efficiently simulate any discrete quantum computation. Thus, in this general
sense, AQC is equivalent to discrete quantum computation.

Adiabatic quantum computation is particularly interesting because of indications
that it is more resilient to decoherence and implementation errors than the
discrete model.
While most schemes for implementing AQC oracles involve approximating them
by a sequence of discrete unitary gates~\cite{Ali,DamVazirani}, robustness of 
potential implementations of the time-dependent Hamiltonians has also been 
studied. For example, Childs et al.~\cite{Robust} considered errors due to 
environmental decoherence and imperfect implementation in the latter approach. 
If the time dependent algorithm Hamiltonian is $H(t)$, they considered 
the actual Hamiltonian to be $H(t)+K(t)$, where $K(t)$ is an error Hamiltonian. 
In particular, $K(t)$ can be a perturbation in the final Hamiltonian.
Through numerical simulations, they demonstrated robustness of AQC for small 
instances of combinatorial search problems against such errors.
\AA berg et al.~\cite{Searchdecoher2,Searchdecoher} investigated robustness to decoherence in the instantaneous
eigenvectors for local~\cite{Searchdecoher} and global adiabatic quantum search~\cite{Searchdecoher2}.
They showed that as long as Hamiltonian dynamics is present, asymptotic 
time complexity is preserved in both local and global cases.
However, in case of pure decoherence, the time complexity of local search
climbs to that of classical. For global evolution, it becomes costlier than
classical: $N^\frac{3}{2}$, where $N$ is the size of the list.

In this paper, we show how to use simple results from linear algebra 
to obtain bounds on the running time of algorithms that obey the adiabaticity
condition locally. 
The technique is useful when the gap between the two lowest eigenvalues is not 
known at every instant of the evolution.
As an illustration and running example, we investigate the behaviour of the eigenvalue spectrum when
the final oracle Hamiltonian for the unordered search problem is perturbed in all non-zero 
elements by an amount at most polynomial in $\log N$. 

In the unperturbed case, the gap between the lowest two eigenvalues is specified at every instant by a nice 
closed form expression~\cite{SearchLocalAdiabatic,DamVazirani}. This makes it rather easy to apply local 
adiabaticity. Perturbation deprives us of this facility. 
We show a work-around
by lower-bounding the gap between the two lowest eigenvalue curves with straight
lines whose slopes are obtained using the Wielandt-Hoffman theorem from
the eigenvalue perturbation theory of symmetric matrices.
The schedule can then be adjusted to satisfy the adiabaticity condition locally.
The bound obtained by our method
is commensurate with existing results--we obtain only a polynomial
speed-up over the global algorithm~\cite{com2,com1,comZ}. 
Tighter bounds on eigenvalue perturbations
will possibly yield Grover speed-up, indicating the resilience of the
adiabatic algorithm to perturbations in the final Hamiltonian.
However, we believe that the techniques that we introduce in this paper can be applied in 
other AQC settings as well.

The paper is arranged as follows. The next section gives a brief discussion of the
adiabatic quantum computing paradigm.
Section 3 discusses the adiabatic search problem and its perturbed version. 
In section 4.1  we show that for the Hamiltonian in question, there exists a non-zero
gap between the two lowest eigenvalues at all times.  The proof largely follows that of Rao~\cite{MV},
simplified for the present case. However the minimum gap turns out to be exponentially small,
and therefore a global schedule is of limited value. In section 4.2 we show our 
method to obtain a local schedule that provides a polynomial speed-up.
Section 5 concludes the paper.
\section{Preliminaries}
In this section we give a brief overview of the adiabatic gap theorem and
its application to quantum computing.
For details the reader is referred to the text by Messiah~\cite{Messiah}.
Let $H(s)$, ($0\leq s\leq 1$), be a time dependent single-parameter Hamiltonian 
for a 
system having an $N$ dimensional Hilbert space.
 Let the eigenstates of $H(s)$ be given by $| l;s\rangle$ and the eigenvalues by $ \lambda_l(s)$,
 with $\lambda_0(s) \leq \lambda_1(s) \leq \ldots \leq \lambda_{N-1}(s)$ for $0\leq s \leq 1$. 
Suppose we start with the initial state of the system $|\psi (0)\rangle$ as $| 0;0 \rangle$ (the ground state of $H(0)$)
  and apply the Hamiltonian $H(s)$, $0\leq s\leq1$, to evolve it to $|\psi (1) \rangle$ at $s=1$. Then the quantum adiabatic theorem states that for a ``large enough'' delay, the final state of the system $| \psi(1)\rangle $  will be arbitrarily
close to the ground state $| 0;1 \rangle$  of $H(1)$.
Specifically,
$|\langle 0;1|\psi(1)\rangle|^2\rightarrow 1$ if the delay schedule $\tau(s)$
satisfies the adiabaticity condition at every $s$:
\begin{equation}
  \tau(s) \gg \frac{  || \frac{d}{ds} H(s)  ||_2 } {g(s)^2},
\end{equation}
where $g(s)$ is the gap between the two lowest eigenvalues at $s$.
In case $g(s)$ is difficult to determine for every $s$, which is the case with most Hamiltonians, we impose a 
more conservative \emph{global} delay schedule for the entire duration $T$ of the evolution:
\begin{equation}
  T \gg \frac{ \max_{0\leq s \leq 1}  || \frac{d}{ds} H(s)  ||_2 } {g^2_{min}},
\end{equation}
where $g_{min} = \min_{0\leq s \leq 1} ( \lambda_1(s) - \lambda_0(s) )$.
However, if we do have an estimate of $g(s)$ for every $s$, we can use a 
varying delay schedule
that satisfies the adiabaticity condition \emph{locally} at every instant $0\leq s \leq 1$.
Then we can reduce the running time to
\begin{equation}
\int_{s=0}^1\frac{||\frac{d}{ds} H(s)  ||_2 ds}{g(s)^2}.
\end{equation}
If $H(s)$ is a polynomial-sized linear interpolation, $||\frac{d}{ds} H(s)||_2$ is a polynomial-sized 
quantity independent of $s$ and the running time is of the order 
$\int_{s=0}^1\frac{ds}{g(s)^2}.$

\section{Perturbed Unordered Search}
Consider a list of $N=2^n$ elements, say bit-strings from 
$\{0,1\}^n$. 
The unordered search problem may be stated as follows.
Given a function $g:\{0,1\}^n\rightarrow\{0,1\}$ such that $g(u)=0$ for
a special element $u$ and $1$ for all others, find $u$. While any classical
algorithm requires $O(N)$ queries to $g$, Grover's celebrated ``discrete"
quantum algorithm accomplishes the search in $O(\sqrt{N})$ queries only~\cite{Grover}.
A similar speed-up was demonstrated by van Dam et al.~\cite{DamVazirani}
and Roland and Cerf~\cite{SearchLocalAdiabatic} for this problem with AQC, which 
we discuss now.
To begin with, note that the initial Hamiltonian $H(0)$ should be independent of the solution, with the 
restriction that it should not be diagonal in the computational basis 
\cite{DamVazirani}. Moreover, the ground state of the initial Hamiltonian  should
be a uniform superposition of all candidate solutions and easy to prepare.
The following Hamiltonian satisfies the above conditions for searching an unordered list~\cite{DamVazirani,SearchLocalAdiabatic}:
 \begin{equation}
 H(0) = \sum_{z \in \{0,1\}^n \backslash 0^n} | \hat{z} \rangle \langle \hat{z} |,
 \end{equation}
 where each $| \hat{z} \rangle$ is a basis vector in the `Hadamard' basis given by
\begin{displaymath}
{ | \hat{z} \rangle } =
\frac{1}{\sqrt{2^n}}\left ( \begin{array}{cc}
1  & 1 \\
1 & -1 \\
\end{array}  \right)^{\otimes n} | z \rangle
\end{displaymath}
and the ground state is $\frac{1}{\sqrt{2^n}} \sum_{z \in \{0,1\}^n} | z \rangle$, which is easy to construct.
The final Hamiltonian, then, is $\sum_{z \in \{0,1\}^n \backslash u} | z \rangle \langle z |$.

In this paper, we consider the case when the final Hamiltonian is perturbed in the
non-zero entries. 
In other words, the final Hamiltonian is given by
\[
  H(1) = \sum_{z \in \{0,1\}^n} f(z) | z \rangle \langle z |,
\]
 where $\{| z\rangle\}$ form the ``computational basis" of the Hilbert space of the system and
 $f:\{0,1\}^n\rightarrow\mathbb{R}_{poly(n)}$ is a function that behaves 
as follows. $f(u)=0$ for a special element $u\in\{0,1\}^n$. For
all other elements $z$, $f(z)>0$. The problem is to minimize $f$, that is,
to find $u$.

Given the initial and final Hamiltonians, define the interpolating Hamiltonian as
\begin{equation}
H(s) = (1-s) H(0) +s H(1).
\end{equation}
We start in the ground state of $H(0)$ and evolve to $H(1)$ slowly enough and end up in its ground state. Since $f(z)$ is bounded by a polynomial in $n$ for all $z\in\{0,1\}^n$,  
so is $|| \frac{d}{ds} H(s)  ||_2 $. The factor deciding the running time
is therefore the denominator $g^2_{min}$. 

\section{The Bounds}
In what follows, we will assume $f(z)\neq f(z')$ for $z\neq z'$.
The motivation for this assumption is two-fold: first, it makes for a neater presentation
of our technique; secondly, given the random nature of noise and implementational error,
it is reasonable to assume that no two non-zero elements of the diagonal will be perturbed by the same amount.
For the case when there do exist $z\neq z'$ such that $f(z)=f(z')$, the subsequent
discussion requires only minor modification.

\subsection{Global Evolution}
By the nature of $H(s)$, the characteristic equation is independent of the permutations of
the diagonal elements of the final Hamiltonian. Therefore, without loss of 
generality, we will follow the convention that
$f(u)=0<f(z_1)<\ldots < f(z_{N-1})$.
First, we make sure that the gap between the two lowest eigenvalues is non-zero at all times.
For that, we evaluate the characteristic equation of $H(s)$, in much the same way as in lemma 1 of~\cite{MV}, and~\cite{comZ}. 
\begin{lemma} \label{lemma1}
The characteristic equation of $H(s)$ is 
\newline \noindent
$c(\lambda) =(1-s-\lambda) \Bigg[\prod_{k=1}^{N-1}(1-s+sf(z_k) -\lambda\big)
-\frac{1-s}{N}\sum_{j=1}^{N-1}\frac{ \prod_{k=1}^{N-1}
\big(1-s+sf(z_k) -\lambda\big)}{ 1-s + sf(z_j) -\lambda}\Bigg]
-  \frac{1-s}{N}\prod_{k=1}^{N-1}\big(1-s+sf(z_k) -\lambda\big)
=0.$
\end{lemma}
\begin{proof}
To evaluate the eigenvalue curves, we evaluate 
$ | H(s) - \lambda I | = 0$.
Subtracting the last column of this determinant from all other columns and using $x_0$ for $1-s-\lambda$, 
 $x_1$ for $1-s + sf(z_1) - \lambda$ and so on, up to $x_{N-1}$ for $1-s + sf(z_{N-1}) -\lambda$, we have
\begin{displaymath} 
\left | \begin{array}{ccccccc}
x_0 & 0 & 0 & \ldots & &\ldots & -\frac{(1-s)}{N} \\
0 & x_1 & 0 &\ldots & & \ldots&-\frac{(1-s)}{N} \\
\vdots & 0 & \ddots & \vdots & & x_{\frac{N}{2} -1} &\vdots \\
-x_{N -1} & -x_{N-1} & \ldots & &  &-x_{N -1} & (x_{N-1} -\frac{(1-s)}{N}) \\
\end{array}  \right |_{N\times N} 
\end{displaymath}
\indent \indent  \indent  \indent  \indent  \indent  \indent \indent \indent \indent  $=0.$

Expanding this determinant gives the required characteristic equation.
$\square$
\end{proof}


The following lemma is reminiscent of lemma 2 of~\cite{MV}, and~\cite{comZ}. We provide a somewhat simpler proof for the present case.

\begin{lemma} \label{lemma2}There exists exactly one root (i.e. an eigenvalue curve) of the characteristic equation in the intervals
$(0,1-s)$, $(1-s, 1-s + sf(z_1))$, $\ldots$, $(1-s + sf(z_{N-2}), 1-s +sf(z_{N-1}) )$; for $0<s<1$.
\end{lemma}
\begin{proof}
For any interval, let the $c_l(\lambda)$ and $c_u(\lambda)$ denote the value
of the characteristic polynomial at the lower and upper boundary respectively.

We analyze the problem as three cases:

(1) The interval $(0,1-s)$: 

In this case, $c_l(\lambda) =(1-s) \Bigg[\prod_{k=1}^{N-1}(1-s+sf(z_k))
-\frac{1-s}{N}\sum_{j=1}^{N-1}\frac{ \prod_{k=1}^{N-1}
\big(1-s+sf(z_k)\big)}{ 1-s + sf(z_j)}\Bigg]
-  \frac{1-s}{N}\prod_{k=1}^{N-1}\big(1-s+sf(z_k)\big)$.
This is a positive quantity for $s\in (0,1)$, as can easily be 
verified.
Moreover,
$c_u(\lambda) = -\frac{1-s}{N} \prod_{k=1}^{N-1}sf(z_k)$ is negative. Therefore,
there exists a root in the interval $(0,1-s)$ for $0<s<1$.

(2) The interval $(1-s, 1-s + sf(z_1))$:

Clearly, $c_l(\lambda)$ for this interval is same as $c_u(\lambda)$
of the previous case, which is negative.
But $c_u(\lambda)= -sf(z_1)\Bigg[-\frac{1-s}{N}\prod_{k=2}^{N-1}s(f(z_k)-f(z_1))\Bigg]$ is positive.
Thus, there exists at least one root in this interval also.

(3) The intervals $(1-s + sf(z_{i}), 1-s +sf(z_{i+1}) )$, $1\leq i\leq N-2$:

The values of the characteristic polynomial at the boundaries are
respectively

$c_l(\lambda)= -sf(z_i)\Bigg[-\frac{1-s}{N}
\prod_{k=1}^{i-1}s(f(z_k)-f(z_i))\prod_{k=i+1}^{N-1}s(f(z_k)-f(z_i))\Bigg]$
and

$c_u(\lambda)= -sf(z_{i+1})\Bigg[-\frac{1-s}{N}
\prod_{k=1}^{i}s(f(z_k)-f(z_{i+1}))\prod_{k=i+2}^{N-1}s(f(z_k)-f(z_{i+1}))\Bigg]$.

Notice that in either case, every element in 
the first product is negative and that in the second is positive. This
is because $f(z_k)<f(z_i)$ for $1\leq k \leq i-1$ and $f(z_k)>f(z_i)$ for 
$i+1\leq k \leq N-1$. Therefore, the over-all sign is decided by the number
of elements in the first product only.
But the number of such elements in $c_l(\lambda)$ and $c_l(\lambda)$ differ
by one. Thus $c_l(\lambda)$ and $c_u(\lambda)$ differ in sign.
Therefore, there exists a root in each of these intervals.

Hence, there is a root in each interval.
Given that (i) there are $N$ open intervals bounded by $N$ straight line eigenvalue curves, 
(ii)  there is at least one root in each interval and (iii) there are $N$ roots in all, the lemma follows.
$\square$
\end{proof}

\begin{figure}
\begin{center}
\includegraphics[width=0.7\textwidth]{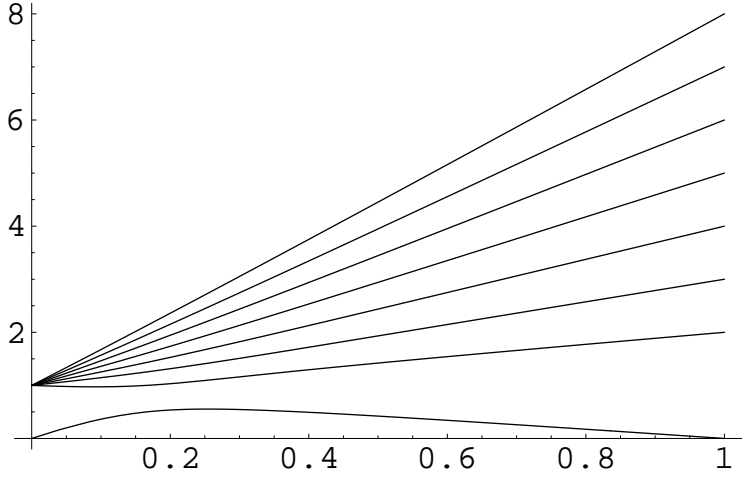}
\caption{Example Eigenvalue Curves}
\end{center}
\end{figure}

By virtue of the above lemma, we can speak of one curve being ``above" another. We label the curves
as $\lambda_0(s),\lambda_1(s),\ldots,\lambda_{N-1}(s)$, starting from below.
Thus, $\lambda_0(s)$ lies between the lines $\lambda(s)=0$ and $\lambda_1(s)$.
See figure 1 for example. 

As a consequence of the above lemma, we are guaranteed a non-zero gap 
between $\lambda_0(s)$ and $\lambda_1(s)$ for all $s\in[0,1]$.

What about the minimum gap between $\lambda_0(s)$ and $\lambda_1(s)$?
It turns out that this gap is exponentially small~\cite{comZ}.
Making use of the properties of the present problem, we 
arrive at this conclusion in a different manner.

We first discuss the case when $f(z_1)$ is large, say greater than $1$. 
Consider the characteristic polynomial
$c(\lambda) =(1-s-\lambda) \Bigg[\prod_{k=1}^{N-1}(1-s+sf(z_k) -\lambda\big)
-\frac{1-s}{N}\sum_{j=1}^{N-1}\frac{ \prod_{k=1}^{N-1}
\big(1-s+sf(z_k) -\lambda\big)}{ 1-s + sf(z_j) -\lambda}\Bigg]
-  \frac{1-s}{N}\prod_{k=1}^{N-1}\big(1-s+sf(z_k) -\lambda\big)
.$

By lemma 2, the line $\lambda(s) = 1-s$ separates $\lambda_0$ and $\lambda_1$.
Thus, if 

(i) there exists an 
$s_0$ where the above polynomial is divisible by two ``$(1-s_0-\lambda)$ 
factors" and 

(ii) all other curves ($\lambda_2(s)$ through $\lambda_{N-1}(s)$) are at 
least an inverse polynomial distance above the point $(s_0,1-s_0)$, 

it would imply that $\lambda_0(s_0)$ and $\lambda_1(s_0)$ are exponentially close to the line 
$\lambda(s)=1-s$, and in turn, to each other.

Clearly, points very close to $s=0$ are not candidates, as there are too many
curves in the vicinity. Points close to $s=1$ are also ruled out, as there
is only one $(1-s-\lambda)$ factor, corresponding to the curve $\lambda_0(s)$:
$\lambda_1(s)$ is $f(z_1)$ units above.

Consider a candidate point $s_0=\frac{1}{p(n)}$ where $p(n)$ is a polynomial in $n$ 
such that
$p(n)>max_{k}(f(z_k))$.
At $s_0=\frac{1}{p(n)}$, $\lambda_2> 1+ \frac{f(z_1)-1}{p(n)}$.
Thus, by lemma 2, all curves above $\lambda_2(s)$ are at least an inverse 
polynomial distance above  $1-s_0= 1- \frac{1}{p(n)}$.

Notice that the polynomial can be divided by one $(1-s-\lambda)$ factor
at an $s$ such that $\frac{1-s}{N}\prod_{k=1}^{N-1}\big(1-s+ sf(z_k) -\lambda\big)\rightarrow 0$. 
Let us test the candidate point $s_0$.
Substituting $1-s$ for $\lambda$ and $\frac{1}{p(n)}$ for $s$, we get
$\frac{p(n)-1}{Np(n)} \prod_{k=1}^{N-1} \big(\frac{f(z_k)}{p(n)} \big)$, which indeed tends to zero.
Therefore, the existence of  one $(1-s-\lambda)$ factor at $s_0$ has been established. Let us now see if there
exists another.
On dividing the polynomial by $(1-s_0-\lambda)$ at $s_0=\frac{1}{p(n)}$, we obtain
$c(\lambda) =\Bigg[\prod_{k=1}^{N-1}\frac{f(z_k)}{p(n)}
-\frac{p(n)-1}{Np(n)}\sum_{j=1}^{N-1}\frac{ \prod_{k=1}^{N-1}
\frac{f(z_k)}{p(n)}}{ \frac{f(z_j)}{p(n)}}\Bigg]
,$
which tends to zero. 

Suppose now that $f(z_1)$
is ``small". If $f(z_1)$ is a constant independent
of $n$ or even $ O(\frac{1}{poly(n)})$,
the argument given above holds exactly. 
However, if $f(z_1)$ is greater than zero by only an exponentially small quantity,
the two $(1-s-\lambda)$ factors that we obtained in the above discussion 
could correspond to $\lambda_2(s)$, thus leaving the possibility of $\lambda_0(s)$
being at a larger distance below $\lambda(s)=1-s $. 
But this is not the case.
Since by assumption  $\lambda_2(s)$ is 
exponentially close to the line $\lambda(s)=1-s$, it can be factored out from
the polynomial, leaving two factors that correspond to $\lambda_0(s)$ 
and $\lambda_1(s)$.

Thus, it turns out that the minimum gap is too small to yield any
significant speed-up for a constant rate global adiabatic algorithm.

\subsection{Local Evolution}
We will now obtain an upper bound on $\int_{s=0}^1\frac{d(s)}{g(s)^2}$ where
$g(s)= \lambda_1(s)-\lambda_0(s)$, for local evolution. 
To that end, the following result
from the perturbation theory of real symmetric matrices will be useful.

\begin{theorem} \label{theorem2} {Wielandt-Hoffman Theorem (WHT):} Let $A$ and $E$ be real symmetric matrices and let $\hat{A}= A+E$. Let the eigenvalues
of A be 
$ \lambda_0 \leq \lambda_1 \leq \ldots \leq \lambda_{N-1} $
and those of $\hat{A}$ be 
$ \hat{\lambda}_0 \leq \hat{\lambda}_1 \leq \ldots \leq \hat{\lambda}_{N-1} $.

\noindent Then
\begin{equation} \label{whteq}
\sum_{j=0}^{N-1}( \lambda_j - \hat{\lambda}_j )^2 \leq \sum_{i=0}^{N-1}\sum_{j=0}^{N-1} | E_{ij} |^2.
\end{equation}$\square$
\end{theorem}
Let $E$ of WHT be  the `perturbation matrix' $H(s+ds) -H(s)$. This provides us with
the r.h.s. of (\ref{whteq}):
\[
\Big(\frac{N-1}{N}\Big)^2(ds)^2 +\frac{N-1}{N} (ds)^2 +\sum_{k=1}^{N-1}\Big(f(z_k) -\frac{N-1}{N}\Big)^2(ds)^2.
\]
For the l.h.s., note that $\lambda_1(s)\ldots\lambda_{N-1}(s)$
are $N-1$ non-crossing curves packed inside a polynomially bounded
gap and each curve is bounded by two straight lines by lemma 2.

Thus, there are at most a polynomial $q(n)$ number of the gaps $f(z_i)-f(z_{i-1})$ 
that are at least inverse polynomially wide. 
All other curves are packed between the
straight lines $\lambda = (f(z_i)-1)s +1$ and $\lambda = (f(z_{i-1})-1)s +1$
where $f(z_i)-f(z_{i-1})=O(\frac{n^c}{N})$ for a constant $c$. Therefore
the slopes of these curves can be approximated by the slopes of one of the 
enclosing lines. Hence, the l.h.s. is
\[
d\lambda_0^2 + d\lambda_{i_1}^2 +\ldots + d\lambda_{i_{q(n)}}^2
+ \sum_{k\in\{1,\ldots,N-1\}\backslash\{i_1,\ldots,i_{q(n)}\}} (f(z_k)-1)^2 (ds)^2
\]
Substituting in~\ref{whteq}, we get an upper bound on
$d\lambda_0^2 + d\lambda_{i_1}^2 +\ldots + d\lambda_{i_{q(n)}}^2$. In particular,
this is also an upper bound on $\frac{d\lambda_0}{ds}$ and $\frac{d\lambda_1}{ds}$.\footnote{A similar 
bound can be obtained using
a general result of Ambainis and Regev~(\cite{AmbReg}, Lemma 4.1).
Nevertheless, we presented a different way
to demonstrate the possibility of problem specific approaches which can yield
tighter bounds than the general one.}
This can be used to obtain an upper bound on the total time required, as we will see soon.

We illustrate the technique for the case when $q(n)=1$ and $f(z_1)\geq1$.

\begin{lemma}
If $q(n)=1$ and $f(z_1)\geq1$,
\begin{equation} \label{slope}
\bigg( \frac{d\lambda_0(s)}{ds} \bigg)^2 + \bigg( \frac{d\lambda_1(s)}{ds} \bigg)^2  \leq (f(z_{N-1})-1)^2 +2n + \frac{2}{N}\sum_{k=1}^{N-2}f(z_k).
\end{equation}
\end{lemma}

\begin{proof}
By the preceding argument, the l.h.s. of~\ref{whteq} is given by

$\sum_{k=1}^{N-2} (f(z_k)-1)^2(ds)^2 + (d\lambda_0(s))^2 +(d\lambda_1(s))^2$.
Substituting in equation~\ref{whteq}, we get

\noindent$\sum_{k=1}^{N-2} (f(z_k)-1)^2(ds)^2 + (d\lambda_0(s))^2 +(d\lambda_1(s))^2$ 

$\leq (\frac{N-1}{N})^2(ds)^2 +\frac{N-1}{N} (ds)^2 +\sum_{k=1}^{N-1}(f(z_k) -\frac{N-1}{N})^2(ds)^2$.

\noindent Rearranging some terms we get,

\noindent$(\frac{d\lambda_0}{ds})^2 +(\frac{d\lambda_1}{ds})^2
\leq (\frac{N-1}{N})^2 +\frac{N-1}{N} + (f(z_{N-1}) -\frac{N-1}{N})^2
$

$+\sum_{k=1}^{N-2}\big( (f(z_k) -\frac{N-1}{N})^2 - (f(z_k)s -\frac{N-1}{N})^2
\big)^2$.

\noindent Or,

\noindent$(\frac{d\lambda_0}{ds})^2 +(\frac{d\lambda_1}{ds})^2
\leq (\frac{N-1}{N})^2 +\frac{N-1}{N} + (f(z_{N-1})s -\frac{N-1}{N})^2
+ \frac{1}{N}\sum_{k=1}^{N-2}(2n+2f(z_k) -\frac{2N-1}{N})$.

Simplifying and ignoring small terms, we get
\[
\Big(\frac{d\lambda_0}{ds}\Big)^2 +\Big(\frac{d\lambda_1}{ds}\Big)^2
\leq (f(z_{N-1})-1)^2 +2n + \frac{2}{N}\sum_{k=1}^{N-2}f(z_k).
\]
$\square$
\end{proof}

The lemma implies that $|\frac{d\lambda_0}{ds}|, |\frac{d\lambda_1}{ds}| \leq 
\sqrt{(f(z_{N-1})-1)^2 +2n + \frac{2}{N}\sum_{k=1}^{N-2}f(z_k)}$.

Using this, we will estimate $g(s)=\lambda_1(s)-\lambda_0(s)$. We conservatively 
approximate $\lambda_0(s)$ and $\lambda_1(s)$ by straight lines to get 
a lower bound on $g(s)=\lambda_1(s) -\lambda_0(s)$. 

We divide the interval $[0,1]$ into three parts 
$[0,a]$, $[a,b]$ and $[b,1]$, corresponding to the parts when
$\lambda_0$ approaches $\lambda=1-s$, when both $\lambda_0$ and $\lambda_1$ are
close to $\lambda=1-s$, and when $\lambda_1(s)$ rises
away from $\lambda=1-s$ respectively.

Consider the first interval. Suppose $l^1_0(s)$ and $l^1_1(s)$ are lines
such that $l^1_0(s)\geq \lambda_0(s)$ and $l^1_1(s)\leq\lambda_1(s)$.
Then the gap between these lines is a lower bound on $\lambda_1(s)-\lambda_0(s)$
in the interval $(0,a)$.

Denote $\sqrt{(f(z_{N-1})-1)^2 +2n + \frac{2}{N}\sum_{k=1}^{N-2}f(z_k)-1}$
by $m$.
We take $l^1_0(s)=ms$ and $l^1_1(s)=1-s$. 
Thus, the gap in the first interval is given by $g_1(s)\geq 1-s-ms$.
Similarly, for the third interval, we choose $l^3_0(s)=1-s$ and
$l^3_1(s)= ms - m +f(z_1)$.
Therefore, the gap $g_3(s)$ in this interval is greater than
$ms - m +f(z_1)-(1-s)$. We will conservatively take the gap in the entire second interval to be $g_{min}$. See figure 2 for a rough sketch.

\begin{figure}
\begin{center}
\includegraphics[width=0.6\textwidth]{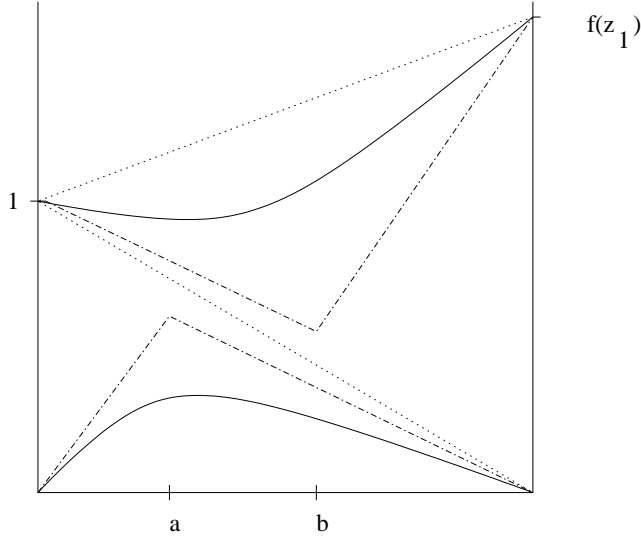}
\caption{Bold lines show actual eigenvalue curves and the dashed lines show
lines that are assumed to bound the gap.}
\end{center}
\end{figure}
Let $a$ be specified by a point on $l^1_0(s) = ms$  that lies vertically below
the line $\lambda=1-s$ by a distance $g_{min}$.
Thus, $a=\frac{1-g_{min}}{m+1}$. Similarly, $b$ is specified by a
point on the line $l^3_1(s)=ms - m +f(z_1)$ that is above $\lambda=1-s$ by
$g_{min}$. Then, $b=\frac{m+1-f(z_1)+g_{min}}{m+1}$.

For the gap $g_2(s)$, recall that one of $\lambda_0$ and $\lambda_1$
is away from the line $1-s$ by a margin of $g_{min}$.
Since this gives the deviation of only one
of $\lambda_0$ and $\lambda_1$ from $1-s$, this is a conservative estimate of $g_2(s)$.

Therefore the total delay factor is given by
\begin{eqnarray}
T &\gg& \int_0^a\frac{ds}{g_1(s)^2} + \int_a^b \frac{ds}{g_2(s)^2}+\int_b^1 \frac{ds}{g_3(s)^2}\nonumber \\
 &\leq& \int_0^a\frac{ds}{\Big(1-s(1+m)\Big)^2}
    + \frac{1}{g_{min}^2} \int_a^bds 
    + \int_b^1 \frac{ds}{\Big((m+1)(s-1) +f(z_1) \Big)^2} \nonumber \\
    &=& \frac{1}{m+1}\Big(\frac{1}{g_{min}}-1\Big) + \frac{m+2g_{min}-f(z_1)}{g_{min}^2(m+1)} + 
    \frac{1}{m+1}\Big(\frac{1}{g_{min}} -\frac{1}{f(z_1)}\Big) \nonumber\\ \nonumber
    &\approx& \frac{m-f(z_1)}{(m+1)g_{min}^2}.
\end{eqnarray}

Let us summarize:
\begin{theorem} 
Let the non-zero elements of the final Hamiltonian of the local 
adiabatic search algorithm be $f(z_1)\neq f(z_2)\neq f(z_{N-1})$, where
each $f(z_i)$ is of size at most  $O(\log N)$. 
Then, the time taken to evolve to the solution state of the final
Hamiltonian is 
$O(D\frac{m-f(z_1)}{(m+1)g_{min}^2})$, where 
$m=\sqrt{(f(z_{N-1})-1)^2 +2n + \frac{2}{N}\sum_{k=1}^{N-2}f(z_k)-1}$,
$g_{min} = \min_{0\leq s \leq 1} ( \lambda_1(s) - \lambda_0(s) )$, and
$D=||\frac{d}{ds} H(s)  ||_2$.
\end{theorem}

To obtain tighter upper bounds on the running time, we would require the 
interval $b-a$ to be shorter: short enough to balance the denominator $g_{min}^2$ 
in the second term. This in turn requires tighter bounds on 
$\frac{d\lambda_0}{ds}$ and $\frac{d\lambda_1}{ds}$.

\section{Conclusions}
We introduced a technique for obtaining upper bounds on the running time
of adiabatic quantum algorithms if the eigenvalue spectrum behaves in certain
ways.
We used this technique to investigate the robustness of the adiabatic quantum algorithm for unordered
search when the final Hamiltonian is perturbed in the non-zero entries.
Interesting open problems include tightening
of the bounds, and application of our technique to other quantum adiabatic
algorithms.
\bibliographystyle{plain}
\bibliography{../../thesis/bib}
\end{document}